\documentclass[twocolumn,groupedaddress]{revtex4}
\usepackage{graphicx}

\begin{document}

\title{Theoretical search for superconductivity in Sc$_3X$B perovskites
and weak ferromagnetism in Sc$_3X$ ($X$~=~Tl,~In,~Ga,~Al).}

\author{B. Wiendlocha}

\author{J. Tobola}
\email[Corresponding author, email: ]{tobola@ftj.agh.edu.pl}

\author{S. Kaprzyk}
\affiliation{Faculty of Physics and Applied Computer Science,
AGH University of Science and Technology, al. Mickiewicza 30, 30-059 Krakow, Poland}

\date{\today}

\begin{abstract}
A possibility for a new family of intermetallic perovskite superconductors Sc$_3X$B, with
$X$~=~Tl,~In,~Ga and Al, is presented as a result of KKR electronic structure and pseudopotential phonon calculations.
The large values of computed McMillan--Hopfield parameters on scandium suggest appearance of superconductivity in Sc$_3X$B compounds. On the other hand, the possibility of weak itinerant ferromagnetic behavior of Sc$_3X$ systems is indicated by the small magnetic moment on Sc atoms in two cases of $X$~=~Tl and In. Also the electronic structure and resulting superconducting parameters for more realistic case of boron--deficient systems Sc$_3X$B$_x$ are computed using KKR--CPA method,
by replacing boron atom with a vacancy.
The comparison of the calculated McMillan--Hopfield parameters of the Sc$_3X$B series with corresponding values in MgCNi$_3$ and YRh$_3$B superconductors is given, finding the favorable trends for superconductivity.
\end{abstract}

\keywords{superconductivity, weak itinerant ferromagnetism}

\maketitle

\section{\label{intro}Introduction}

The motivation to search for a superconductivity in the intermetallic series of Sc$_3X$B compounds was prompted by the very interesting and quite non--typical superconductivity of MgCNi$_3$ \cite{nature1} perovskite,
with the critical temperature T$_c \simeq 8$~K.
First of all, the large amount of nickel atoms would suggest ferromagnetic properties in this compound and propensity to form  magnetic state is really large, due to the van Hove singularity near the Fermi level ~\cite{pickett1}.
This may be a reason of some unexpected effects, which were observed experimentally in this
material \cite{press1,press2,zbcp,penetr}$^,$~\footnote{For a review of the MgCNi$_3$ properties see also the recent papers on this subject, e.g. Refs.~\cite{isef,role_c-calc}.}, despite MgCNi$_3$ belongs to electron--phonon type superconductors \cite{spheat,nmr} with a very large isotope effect on carbon \cite{isef}.

Among other intermetallic perovskites, similar to MgCNi$_3$, a single compound was
earlier reported to be a superconductor: YRh$_3$B with T$_c$~=~0.76~K \cite{ybrh3}.
However, since discovery of superconductivity in MgCNi$_3$, no other superconductor has been found so far.
Systematic experimental study \cite{schaak} showed some indications of superconductivity in
CaB$_x$Pd$_3$ ($T_c$~$\simeq$~1~K) and NbB$_x$Rh$_3$ ($T_c$~$\simeq$~6~K), but the superconducting phase
was finally not identified. Also, no evidence of superconductivity was found experimentally
in ZnCNi$_3$ \cite{zncni3}, despite the calculated electronic structure beeing very close to that of
MgCNi$_3$. Here the carbon deficiency was proposed as a possible explanation  \cite{zncni3_2}.

The predictions of superconductivity in one of the compounds from the entitled series -- Sc$_3$InB -- was announced in our conference paper \cite{pm05}.
Interesting properties of Sc$_3$InB encouraged us to investigate electronic structure and electron--phonon coupling along the whole isoelectronic series of Sc$_3X$B compounds, with $X$~=~Tl,~In,~Ga,~Al, and detailed results of our theoretical study is presented in this work.

\begin{figure}[b]\label{fig:cryst}
\includegraphics[width=.30\textwidth]{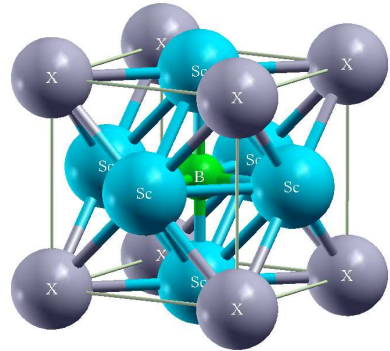}
\caption{ Crystal structure of the Sc$_3X$B perovskite. The atomic positions are: $X$: 1a (0,0,0),
B: 1b ({1/2},{1/2},{1/2}) and Sc: 3c (0,{1/2},{1/2}), ({1/2},0,{1/2}), ({1/2},{1/2},0). If central B atom is removed, perovskite structure changes into Cu$_3$Au--type.}
\end{figure}

The crystal structure of Sc$_3X$B (space group P$m$-$3m$, CaTiO$_3$ type) is given in Fig.~\ref{fig:cryst}.
The synthesis of Tl-- and In--containing cubic
perovskites was reported \cite{holleck}, while
Ga-- and Al--containing compounds are still hypothetical.
The perovskite structure of Sc$_3X$B may also be viewed as a Cu$_3$Au--type cubic Sc$_3X$, with additional boron atom placed in the center (Fig.~\ref{fig:cryst}).
In particular, the Sc$_3$In system was found to exist, in two polymorphic forms \cite{phase_d}: one is the above mentioned, poorly known cubic Cu$_3$Au, detected under a high pressure \cite{cannon}, and the second one is hexagonal Ni$_3$Sn--type, in which weak ferromagnetic properties were observed  \cite{matt}.
The ferromagnetism of hexagonal Sc$_3$In guided us to study possibility of magnetic behaviors in the binary series of cubic Sc$_3X$ compounds.
Indeed, the computations shows
that the cubic Sc$_3$In is magnetic as well as the hexagonal phase.
For other systems the results of electronic structure calculations
are also presented, although we are not aware of the existence of these compounds. Noteworthy,
spin--polarized computations of Sc$_3X$B shows the ground state to be non--magnetic.
We also intend to underline some correlations between tendency to superconductivity in
Sc$_3X$B on one hand and to weak ferromagnetism in the corresponding Sc$_3X$ on the other hand.

%
%
%
%

\section{Calculations}\label{calc}

First--principle calculations of the superconducting parameters, due
to the subtle nature of such phenomenon, are still challenging problem
in the Density--Functional computations. McMillan \cite{mcm} in his fundamental
work showed, that electron--phonon coupling (EPC) calculations may be decoupled into electronic
and phonon contributions, when using a few approximations. The commonly used formula
for the essential parameter for superconductivity -- EPC constant $\lambda$, resulting from generalization of his concept to multi--atomic compounds, is as follows:
\begin{equation}\label{eq:lam}
\lambda = \sum_i \frac{\eta_i}{M_i\langle \omega_i^2 \rangle} = \sum_i \lambda_i,
\end{equation}
where $i$ corresponds to all atoms in the unit cell, with $\eta_i$ being the electronic, and
${M_i\langle \omega_i^2 \rangle}$ lattice contribution to the electron--phonon interaction parameter.

The electronic part of the EPC constant -- McMillan--Hopfield
parameters $\eta_i$ \cite{mcm,hop}, which describe the response of electrons from
Fermi surface to displacements of atoms, are computed within
the Gaspari--Gy\"orffy \cite{gasgy} method, using so--called Rigid--Muffin--Tin Approximation (RMTA). The relevant formula for $\eta$ is  \cite{pick2} ($i$~subscript is dropped):
\begin{equation}\label{eq:eta}
\eta = \sum_l \frac{(2l + 2)n_l(E_F)n_{l+1}(E_F)}{(2l+1)(2l+3)n(E_F)} \left|\int_0^{R_{MT}}r^2 R_l\frac{dV}{dr}R_{l+1} \right|^2,
\end{equation}
where $V(r)$ is the self--consistent spherically--symmetric potential on given atom site, $R_{MT}$ is
the radius of the $i$--th muffin--tin sphere, $R_l(r)$ is a regular solution of the radial Schr\"odinger
equation (normalized to unity inside the $MT$ sphere), and $n_l(E_F)$ is the {\it l}--th partial density of states at the Fermi level ($E_F$) on the site considered.

This method usually involves three main assumptions \cite{nonrmta}: \\
(i) {\it rigid--ion approximation}, in which potential inside the MT sphere
moves rigidly with the ion, and the change in crystal potential, caused by the atom displacement, is given
by the potential gradient; \\ 
(ii) {\it local--vibration approximation} -- more generally $\lambda =\sum_{ij}\lambda_{ij}$, where $(i,j)$
refer to two atoms in the unit cell; in this approximation off--diagonal terms are neglected, i.e.
$\lambda_{ij} = \delta_{ij}\lambda_i$ (Eq.~(\ref{eq:lam})),  \\
(iii) {\it spherical band approximation}, which leads to only dipole transitions ($l\rightarrow$~$l+1$) in Eq.~(\ref{eq:eta}).

It is known that all mentioned assumptions give satisfactory results for
the transition metal elements and cubic--site symmetry. In simple
metals the assumptions (i) and (ii) generally underestimate \cite{nonrmta2}
EPC due to a poorly screened crystal potential,
whereas, the corrections to (iii) are expected
to be small for cubic transition metals  \cite{butler:nb}.

In order to allow for independent calculations of electron and phonon subsystems,
another simplification is necessary. Estimation of
$\langle \omega_i^2 \rangle$ in Eq.~(\ref{eq:lam}), using exclusively phonon density of states $F(\omega)$ instead of full electron--phonon coupling function $\alpha^2F(\omega)$, requires an assumption, that
the electron--phonon interaction is considered to be independent of
phonon frequency~$\omega$, so the electron--phonon interaction factor $\alpha^2(\omega)$
cancels~\footnote{Definition of the $n$--th frequency moment of the Eliashberg coupling function $\alpha^2F(\omega)$ \cite{grimvall,allen}:
$\langle \omega^n \rangle$ = $\int \omega^{n-1} \alpha^2 F(\omega) d\omega / \int
\omega^{-1} \alpha^2 F(\omega) d\omega$. If $\alpha^2(\omega) \simeq const$, $\langle \omega^n \rangle$ $\simeq$ $\int \omega^{n-1}
F(\omega) d\omega / \int \omega^{-1} F(\omega) d\omega$. $\alpha^2(\omega)$ is the electron--phonon interaction coefficient.}
when calculating $\langle \omega_i^2\rangle$ (see e.g. Ref.~\cite{butler:tr}).

The assumption that EPC magnitude does not change with phonon frequency is well fulfilled e.g. in niobium \cite{sav:met}, but
may not be of the same behavior in multi--atom compounds. However,
the RMTA method was successfully used for
analyzing EPC in many superconducting materials, like pure metals
 \cite{met}, binary alloys  \cite{nb-mo}, {\it A}--15 compounds
 \cite{a15}, transition metal carbides  \cite{carbides,carbides2},
borocarbides  \cite{borocarb} or metal--hydrogen system  \cite{pdh}.
Reasonable results (as far as $\lambda$ is concerned) were usually obtained,
even in such unusual superconductor, as MgB$_2$ \cite{mgb2}. This formalism was also helpful in
discussing phonon--based effects in high--temperature
superconductors  \cite{nonrmta,hts_rmp}. Certainly, in order
to better understand the electron--phonon interactions in a
superconductor, one has to perform more elaborated calculations,
using the Eliashberg gap equations (see, for example,
results for MgB$_2$ \cite{mgb22,mgb23}). Nevertheless, such electronic
structure and phonon calculations, together with the simplified
RMTA framework, are very useful and efficient tools of early stage of looking for new superconducting systems.

In line with this methodology, McMillan--Hopfield parameters and phonon frequency moments
were calculated for Sc$_3X$B compounds, and then, using Eq.~(\ref{eq:lam}), EPC parameters $\lambda$ were
deduced.

Electronic structure calculations were performed with the KKR method, which in case of disordered systems (Sec.~\ref{vacancy}) was
implemented together with coherent potential approximation (CPA), as described in details in Ref.~\cite{cpa}.
Additionally, for ordered compounds computations were performed with the full--potential (FP) code, based on the FP--KKR formalism,
widely discussed by many authors  \cite{MRS91}, with technical details shown
e.g. when applying to Si \cite{kellen95}, or to the electric field gradient calculations \cite{akai}.
In our practice extracting bands was done with
the novel quasi--linear algorithm  \cite{stopa}, which allows
for more precise and less time--consuming band structure calculations,
comparing to conventional techniques.
The crystal potential was constructed within the local density approximation
(LDA) and Perdew--Wang \cite{PW} formula was used for the exchange--correlation part.
For all calculations angular momentum cut--off $l_{max}$~=~3 was
set. Highly converged results were obtained for $\sim$120
{\bf k}--points grid in the irreducible part of Brillouin zone (IRBZ), but they were also
checked for convergence using more dense {\bf k}--mesh and $l_{max}$~=~4.
Electronic densities of states were computed using a tetrahedron {\bf k}--space integration
technique (up to 700 tetrahedrons in IRBZ).
In the case of investigated compounds, the differences between the full potential and spherical potential KKR calculations of the density of states (DOS) values at the Fermi level were of order of a few percents.
The justification for using different kinds of approximations, when computing McMillan--Hopfield parameters, was supported by checking the numerical values on changing different input data (like nonrelativistic vs. semirelativistic calculations, type of exchange--correlation potential, radii of muffin--tin spheres), which we also found to differ on the range of a few percentage. Such accuracy we consider satisfactory to explain trends in the number of systems, but for more detailed analysis the fine computation with the full potential will be necessary, especially if the Fermi surface specific features become important.

Phonon calculations were undertaken for realistic approximation
of the phonon part of the electron--phonon coupling constant $\lambda$, i.e. $\langle \omega_i^2 \rangle$
parameter in Eq.~(\ref{eq:lam}).
We used the PWscf package  \cite{pwscf}, where the plane
wave pseudopotential technique and perturbation theory \cite{dfpt} was implemented.
For Sc, B, and Tl atoms ultrasoft pseudopotentials
were employed, for Al, Ga and In norm--conserving pseudopotentials were taken. LDA parameterization of Perdew and Zunger \cite{PZ} was implemented. Plane--wave kinetic energy and charge density cut--offs were set to 30~Ry and 350~Ry, respectively. The Brillouin zone integration smearing technique of Methfessel and Paxton  \cite{m-p} (with parameter $\sigma$~=~0.02~Ry) was used during the calculations. In order to obtain the phonon DOS $F(\omega)$, first the dynamical matrices on (5,5,5)
{\bf q}-point grid were calculated. Then, using the Fourier transformation on the same grid, real--space
interatomic force constants were computed. The final result -- phonon
densities of states -- were obtained from frequencies calculated from force constants on (10,10,10) grid and using the tetrahedron method.
The combined results of the KKR electronic structure study and the phonon DOS calculations were then used to estimate the EPC constant $\lambda$.

We also estimate the superconducting transition temperature $T_c$ from McMillan formula \cite{mcm} in Eq.~(\ref{eq:tc}),
using the modified factor $\langle \omega \rangle/1.2$ \cite{allen,grimvall}:

\begin{equation}\label{eq:tc}
T_c =  \frac{\langle \omega\rangle}{1.2}\exp\left[-\frac{1.04(1+\lambda)}{\lambda-\mu^{\star}(1+0.62\lambda)}\right]
\end{equation}

The absolute value of $T_c$ depends also on Coulomb pseudopotential parameter $\mu^{\star}$, and its influence on $T_c$ magnitude is additionally discussed.

%
%
%
%

\subsection{Predictions of superconductivity in Sc$_3$XB}\label{sc}

Crystal structure and atomic positions of the Sc$_3X$B system are presented in Fig.~\ref{fig:cryst}.
For all investigated ordered Sc$_3X$B compounds, lattice constants were calculated from total energy minimum, as a first step of the PWscf phonon calculations.
Both KKR electronic structure and phonon computations were then performed with these equilibrium
lattice parameters.
Except for $X$~=~Al, semi--relativistic calculations results are presented here.
In order to obtain the McMillan--Hopfield parameters, the following {\it muffin--tin} spheres radii (in the lattice constant $a_0$ unit) were employed:
$R_{Sc}$ = 0.325, $R_{X}$ = 0.36 and $R_{B}$ = 0.174. Eq.~(\ref{eq:eta}) suggests, that
$\eta_i$ parameter may be sensitive to a choice of $R_{MT}$, so we also checked the
influence of the computational geometry on final results. The $R_{MT}$ spheres variation in the range of~15\% changed the results in about~5\%. The difference in obtained results for somehow arbitrary $MT$ geometry will be treated as an computational accuracy of the $\eta_i$ value.

\begin{table}[t]
\caption{Lattice parameters in Sc$_3X$B series, units: 1~Bohr~=~0.5292~\AA.} \label{tab:lat}
\begin{ruledtabular}
\begin{tabular}{lcc}
Compound & $a$ experimental \cite{holleck} & $a_0$ calculated  \\
\hline
Sc$_3$TlB & 8.541  &  8.650  \\
Sc$_3$InB & 8.618  &  8.610  \\
Sc$_3$GaB &  ---   &  8.550  \\
Sc$_3$AlB &  ---   &  8.610  \\
\end{tabular}
\end{ruledtabular}
\end{table}

\begin{figure}[t]
\includegraphics[width=.45\textwidth]{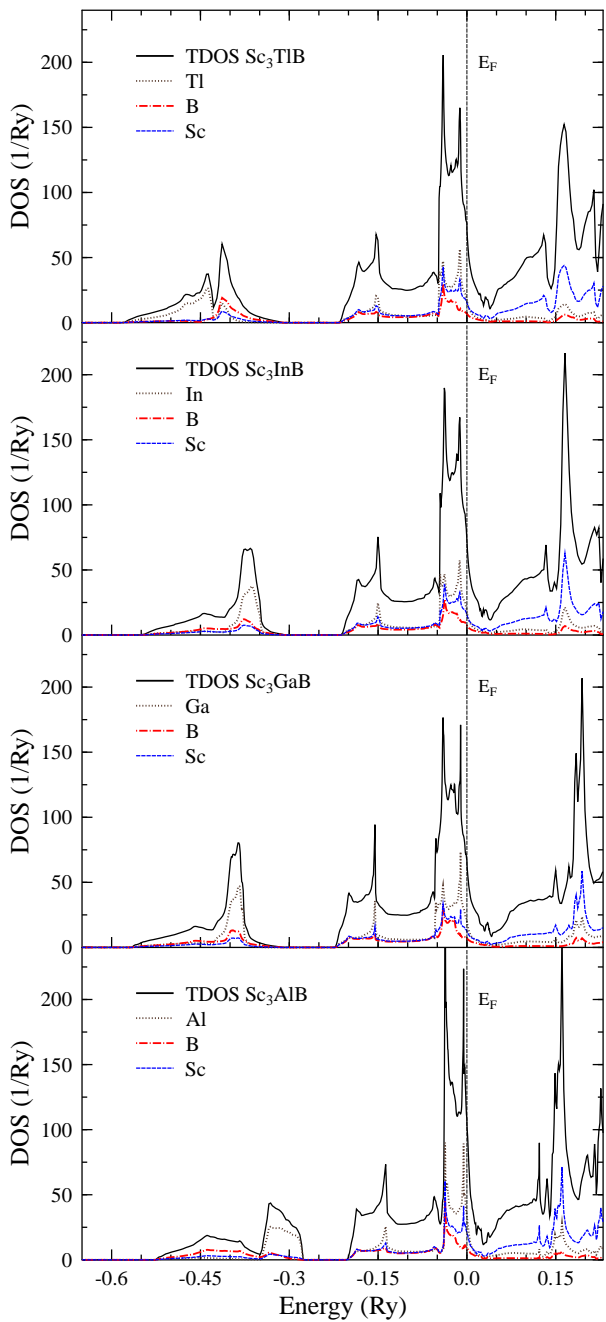}
\caption{\label{fig:xb} KKR total DOS of isoelectronic Sc$_3X$B perovskites calculated at equilibrium lattice constant $a_0$ (Tab.~\ref{tab:lat}).
Site--decomposed densities for $X$, B, Sc are plotted in brown, red and blue, respectively. The Fermi level is shifted to zero and marked by a vertical line.}
\end{figure}

In Table~\ref{tab:lat} results of lattice constant optimization for Sc$_3X$B are summarized. In both existing compounds the agreement between experimental and theoretical values is quite good. The variation of $a_0$ with atom $X$ reflects the change in ionic radius of the $X$ element, where Ga has the smallest radius, and Tl the largest. For Sc$_3$TlB calculated value $a_0$ is larger than the experimental one, which is not so common case.
Similar effect was also observed in YRh$_3$B \cite{ybrh3:el,ybrh3:el2}. The smaller experimental value of $a$ in Sc$_3$TlB could indicate that the measured sample was
boron deficient, as it was earlier suggested for YRh$_3$B \cite{ybrh3:el2}.

Figure~\ref{fig:xb} presents electronic DOS in the Sc$_3X$B compounds.
As we can see, the DOS shape is very similar along this series, due to the same number
of valence electrons.
Noteworthy, even Sc$_3$AlB, where Al contains only $s-$ an $p-$like orbitals, exhibits electronic structure quite similar to other Sc$_3X$B compounds, containing complete $nd^{10}$ shell
from $X$ atom. This shell forms semicore level located about 1~Ry below $E_F$ (not shown).
It supports widely accepted conclusion, that $3d$-Ga, $4d$-In and $5d$-Tl electrons
build core--like levels, and the effect on upper lying electronic states (particularly near $E_F$) is small.

\begin{table}[htb]
\caption{Electronic and phonon properties of Sc$_3X$B. $n(E_F)$ is given in (1/Ry/atom), $\eta$ in (mRy/Bohr$^2$/atom),
$\omega$ in (meV), $M\langle \omega^2 \rangle$ in (mRy/Bohr$^2$).} \label{tab:xb}
\begin{ruledtabular}
\begin{tabular}{lccccc}
Atom & $n(E_F)$ & $\eta_i$ & $\sqrt{\langle \omega_i^2 \rangle}$ & $M_i{\langle \omega_i^2 \rangle}$ & $\lambda_i$\\
\hline
Sc   &  20.0 &  18 &  15.2 &  50   & 0.36  \\
Tl   &  17.4 &   2 &  15.2 &  226  & 0.01  \\
B    &   7.6 &  32 &  61.9 &  204  & 0.16  \\ \hline
Sc   &  20.0 &  19 &  18.5 &  76   & 0.25  \\
In   &  16.3 &   1 &  18.5 &  193  & 0.01  \\
B    &   6.7 &  34 &  58.3 &  181  & 0.19  \\ \hline
Sc   &  19.0 &  18 &  19.0 &  80   & 0.23   \\
Ga   &  18.4 &   2 &  19.0 &  124  & 0.01  \\
B    &  6.5  &  35 &  60.8 &  197  & 0.18  \\  \hline
Sc   &  21.6 &  18 &  21.3 &  100  & 0.18 \\
Al   &  21.5 &   1 &  21.3 &  60   & 0.02  \\
B    &   6.4 &  30 &  58.6 &  183  & 0.16  \\
\end{tabular}
\end{ruledtabular}
\end{table}

\begin{figure}[htb]
\includegraphics[width=.40\textwidth]{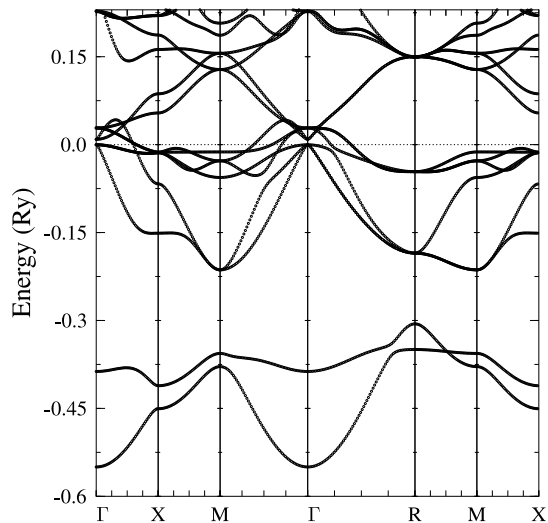}
\caption{\label{fig:bnd}Electronic dispersion curves E({\bf k}) along high symmetry directions in Sc$_3$InB perovskite. The Fermi level is shifted to zero and marked by a horizontal line.}
\end{figure}

The most important part of the density of states near the Fermi level in all compounds is
formed from Sc $d$--states hybridized with $p$--states from B and $X$.
The site--contributions to $n(E_F)$ are presented in Tab.~\ref{tab:xb}~\footnote{The values of $n(E_F)$ presented in Tab.~\ref{tab:xb} for Sc$_3$InB are a bit different from those in Ref.~\cite{pm05}. It comes from applied smaller lattice parameter (the equilibrium value instead of the experimental one) and semi--relativistic calculations. Because the Fermi level is located on the slope of DOS peak, small shift in $E_F$ has noticeable influence on $n(E_F)$. However, $\eta$ values are not infected much, because
they are defined as a \emph{ratio} of resulting densities.}.
Electronic dispersion curves for the representative compound -- Sc$_3$InB -- are shown in Fig.~\ref{fig:bnd}, and they are very similar in other cases (not given).
The DOS sharp peak below $E_F$ comes from a flat band, which is best seen in
E({\bf k}) along the $X$--$M$--$\Gamma$ direction. Generally, the bands in Sc$_3$InB
are more dispersive comparing e.g. to MgCNi$_3$ \cite{dugdale,szajek} or
Sc$_3$In (see~Fig.~\ref{fig:bnd2}). Two separated, lowest lying bands are
formed from $s$--states of $X$ and B atoms, with notably large contribution from Sc.

For our searching, the most interesting feature of electronic structure is related to quite
large McMillan--Hopfield parameters seen along the whole series (see~Tab.~\ref{tab:xb}).
In scandium, typically for transition elements, $d$~--~$f$ scattering channel gives the most
important contribution to $\eta_i$.
For boron and $X$--element only the $p$~--~$d$ channel contributes to $\eta_i$.
Note that although the $\eta_B$ values were found to be the largest in our systems, $\eta_{Sc}$ occurred to be more
important in calculation of $\lambda$, since it is counted three times (3~Sc atoms in unit cell), and has lower value of $M_i\langle \omega_i^2 \rangle$ (see below).
$\eta_{X}$ has negligible value, despite a noticeable density of states at the Fermi level.
The accuracy of computed $\eta_i$, associated with different $MT$ geometry, is about 1~mRy/Bohr$^2$ for scandium and 2~mRy/Bohr$^2$ for boron.

\begin{figure}[t]
\includegraphics[width=.45\textwidth]{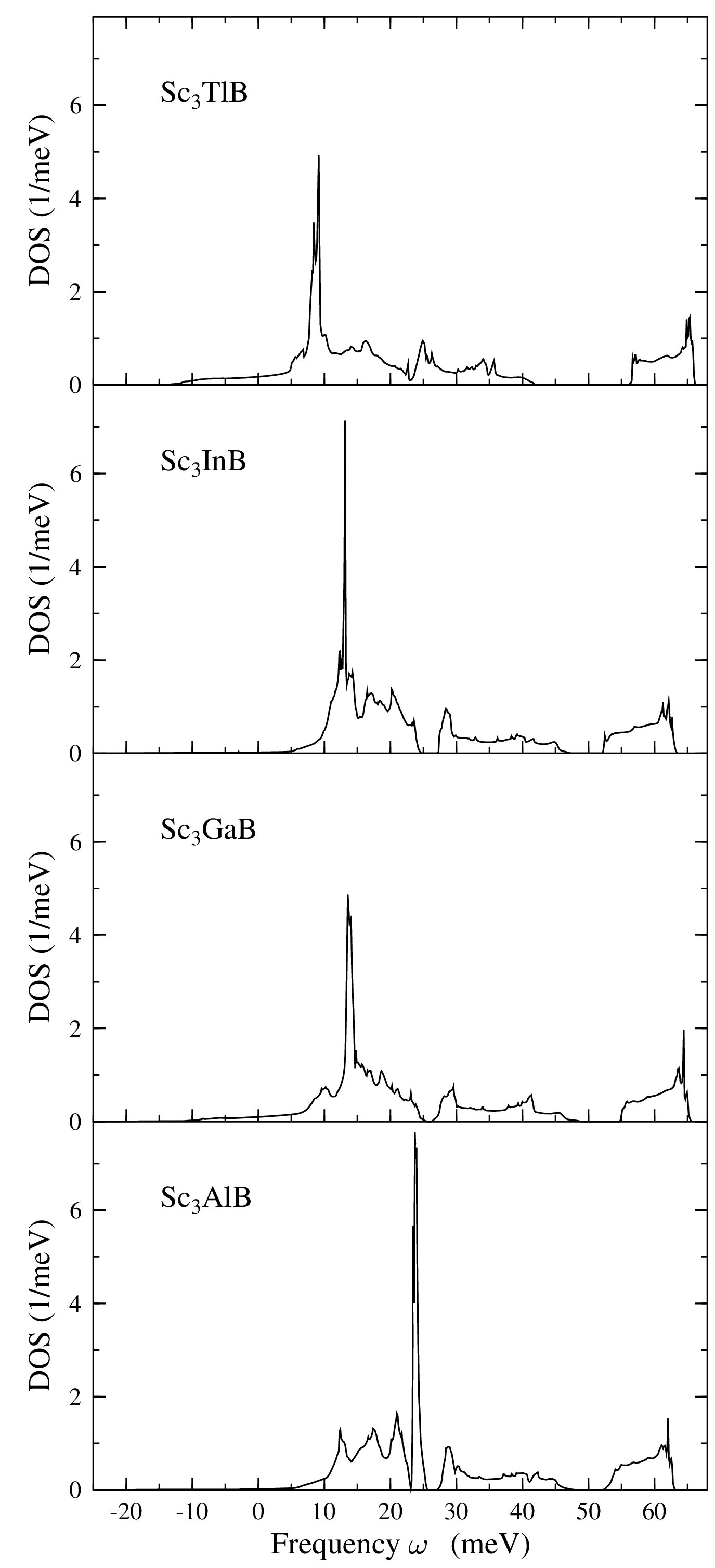}
\caption{\label{fig:ph}Phonon DOS $F(\omega)$ in Sc$_3X$B perovskites.}
\end{figure}

The phonon DOS, and its evolution with the $X$--element, is given in Fig.~\ref{fig:ph}.
The most striking feature of the presented $F(\omega)$ is the 'rigid--band--like' modification with the mass variation of the $X$--element (note that the mass distribution in the unit cell markedly changes).
The highest peak in Fig.~\ref{fig:ph} marks the lowly--dispersive part of acoustic branches, associated with the $X$--atom vibrations. As one can see, when the mass of $X$ is getting smaller, the range of acoustic modes broadens,
which is manifested in a shift of their flat parts towards higher frequencies, i.e. the peak moves from 9~meV ($X$=Tl) to 27~meV ($X$=Al).
Generally, the phonon spectrum consists of two separate areas: the high--frequency part includes essentially
phonon modes of light B, with energies above 50~meV, and the lower--frequency part, below 50~meV, contains
mainly $X$ and Sc states.
This behavior was also characteristic for the dynamic properties of MgCNi$_3$ compound, where carbon vibrations had the highest frequencies \cite{ph}.

Let us remark, that in the phonon calculations for all compounds we met some
problems with imaginary frequencies, appearing near the BZ center ($\Gamma$), and additionally for
$X$~=~Tl and Ga, near $R$ point (in Fig.~\ref{fig:ph} they are visible as negative
frequencies tails). These eigenvalues correspond to three optical modes, and at $\Gamma$ and $R$
points only Sc atoms vibrate in these modes, with eigenvectors lying in three planes perpendicular
to the Sc--B bonds. Physically, the occurrence of imaginary frequencies may indicate either the
instability of the perfect perovskite structure (favoring crystal distortion) or important
anharmonic effects.

Similarly, unstable phonon branches were detected in
MgCNi$_3$ \cite{ph,sav1} occurring at different
high--symmetry points ($X$, $M$). These branches corresponded mainly
to anharmonic Ni vibrations, resulted from a double--well
potential \cite{ph,sav1,exafs},
in which Ni was placed. However, due to the shallowness of the
double well, no stable long--range structural distortion
was found there, and the perovskite structure was stabilized
dynamically. The origin of unstable vibrations in Sc$_3X$B is
attractive problem itself, but more detailed analysis of their dynamical
properties is out of the scope of this paper.

In summary, the influence of the negative frequency range on phonon density
of states is not so critical (the related energy bands are strongly dispersive). This
should not affect much the $\langle \omega_i^2 \rangle$ calculations.
The total weight of the negative frequency area is about 0.5\% in Sc$_3$InB and Sc$_3$AlB, 2.5\%
in Sc$_3$GaB and 5\% in Sc$_3$TlB. So, in the case of $X$~=~In and Al the 'negative tail' was negligible and
cut off in the frequency moments calculations. For two remaining compounds ($X$~=~Ga and Tl) the tail was
cut off at about 3~meV and $F(\omega)$ was extrapolated to reach $\omega=0$ in Debye--like manner.
Because the phonon DOS obtained for Sc$_3$TlB seems less reliable, in this case the evaluation of the superconducting
parameters should be treated rather qualitatively.

Having phonon DOS we may proceed to estimate EPC strength.
The mean phonon frequencies for constituent atoms were computed from the phonon DOS using
the above--mentioned analysis of the spectrum: $\langle \omega_i^2 \rangle$ for $X$ and Sc were calculated
from the lower--frequency part of the spectrum, and were assumed to be the same for both atoms.
As expected, the lowest value of $\langle \omega_i^2 \rangle$ is observed in thallium compound.
The average $\langle \omega_B^2 \rangle$ (for boron) were computed from the high--frequency
part of DOS, and resulted in significantly larger values, due to small mass of B (see~Tab.~\ref{tab:xb}).
Note, that taking the same $\langle \omega_i^2 \rangle$ for all atoms (sometimes practiced), might occur
incorrect \cite{carbides} and disagree with atomic--like character of $\eta_i$
($M_i\omega_i^2$ corresponds to an effective force constant, so it is also a site--dependent quantity).
In systems, where atoms have markedly different masses, the
lightest element would be favored in that way, as shown below.
Therefore, the phonon DOS was divided into high-- and low-- frequency parts
in the  $\langle \omega_i^2 \rangle$ calculations.

\begin{table}[t]
\caption{Total EPC constant $\lambda$ and critical temperature $T_c$ (K) for $\mu^{\star} = 0.13.$ $\langle \omega \rangle$ used in Eq.~(\ref{eq:tc}) is given in (K).} \label{tab:tc}
\begin{ruledtabular}
\begin{tabular}{ccccc}
& Sc$_3$TlB  & Sc$_3$InB & Sc$_3$GaB & Sc$_3$AlB \\ \hline
$\lambda$ & 1.25  & 0.94 & 0.88 & 0.72  \\
$\langle \omega \rangle$ & 182 & 225 & 226 & 263 \\
$T_c$           & 15  & 12 & 10 & 7.5  \\
\end{tabular}
\end{ruledtabular}
\end{table}

The values of total EPC constant $\lambda$, i.e. sum over atomic contributions shown in Tab.~\ref{tab:xb}, are gathered in Tab.~\ref{tab:tc}, together with the critical temperature values. In all investigated compounds the main contribution to $\lambda$ comes from scandium. Due to
the lowest value of $\langle \omega_{Sc}^2 \rangle$ parameter, Sc$_3$TlB has the highest $\lambda$.
Superconducting transition temperature $T_c$ was estimated using Eq.~(\ref{eq:tc}), with typical value of Coulomb pseudopotential $\mu^{\star}$~=~0.13 applied. As one can notice, all compounds with calculated $\lambda$ in the range of 0.7 -- 1.2, and $T_c$~$\sim$~10~K are medium-- or even strong--coupling superconductors in the RMTA framework.

\begin{figure}[htb]
\includegraphics[width=.35\textwidth]{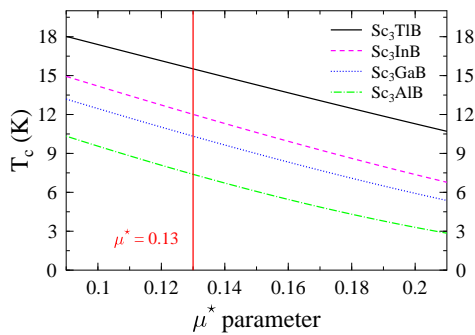}
\caption{\label{fig:tc} Influence of $\mu^{\star}$ parameter on critical temperature $T_c$. Vertical line marks $\mu^{\star}$ = 0.13.}
\end{figure}

To verify the influence of employed $\mu^{\star}$ on $T_c$ magnitude, we plotted the value of $T_c$ versus $\mu^{\star}$ in the reasonable range of 0.09 $< \mu^{\star} <$ 0.21 (Fig.~\ref{fig:tc}). Transition temperatures decrease almost linearly with $\mu^{\star}$ in this range, and even for large value of $\mu^{\star}$ = 0.21, $T_c$ is still enough high to be detectable in the typical low--temperature
measurements.

We also examined how our results changed, when $\langle \omega_i^2 \rangle$ was taken to be equal for all atoms.
For example, in Sc$_3$InB we obtained $\lambda = 1.6$ with $\lambda_B = 1.1$ and $\lambda_{Sc} = 0.5$, when
employed  $\sqrt{\langle \omega^2 \rangle} = 23.4$~meV, as derived from the entire phonon spectrum.
The resulting critical temperature $T_c= 26$~K was much higher than the previous value, and probably much
overestimated, due to the overestimation of boron contribution.

The pressure effect on electronic properties of Sc$_3X$B was additionally inspected by decreasing lattice parameter
in KKR computations (in the range of 0--5\%). The variation of the most important parameters -- total density of states at $E_F$ and McMillan--Hopfield parameters for boron and scandium -- is summarized in Tab.~\ref{tab:press}. The observed tendency is again similar in the whole series. In spite of the $n(E_F)$ decrease with volume shrinking, $\eta_i$ parameters increase with ratio about $d \ln \eta_B / d \ln V$~$\simeq$~$-2$
and $d \ln \eta_{Sc} / d \ln V$~$\simeq$~$-1.5$.

\begin{table}[t]
\caption{Influence of unit cell volume shrinking on electronic critical parameters. \label{tab:press}}
\begin{ruledtabular}
\begin{tabular}{lccc}
Compound & $\frac{dn(E_F)}{dV} $ & $\frac{d\ln \eta_{Sc}}{d\ln V} $ &  $\frac{d\ln \eta_{B}}{d\ln V}$ \\
\hline
Sc$_3$TlB & 0.41   &  -1.4 & -1.9  \\
Sc$_3$InB & 0.41   &  -1.5 & -2.0  \\
Sc$_3$GaB & 0.50   &  -1.2 & -2.3  \\
Sc$_3$AlB & 0.44   &  -1.4 & -2.6  \\
\end{tabular}
\end{ruledtabular}
\end{table}

These results suggest, that external pressure may enhance $T_c$, as observed in MgCNi$_3$
 \cite{press1,press2}, but the opposite effect of lattice stiffening
may be dominant.

\begin{figure}[t]
\includegraphics[width=.40\textwidth]{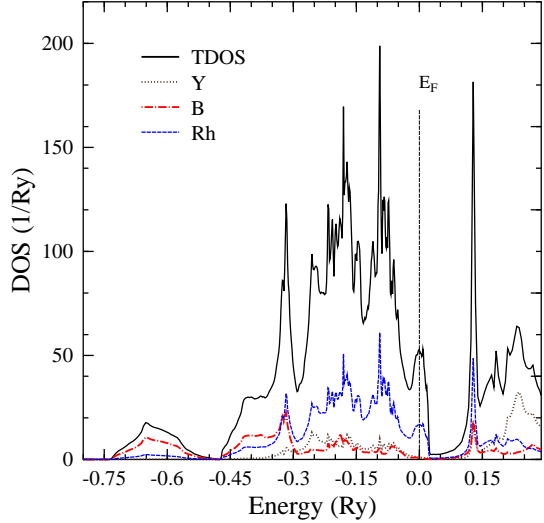}
\caption{\label{fig:ybrh3} KKR DOS of YRh$_3$B calculated at experimental lattice parameter. Site--decomposed densities for Y, B, Rh are plotted in brown, red and blue, respectively. The Fermi level is shifted to zero and marked by a vertical line.}
\end{figure}

To have a better insight into the magnitude of McMillan--Hopfield parameters in Sc$_3X$B, we performed similar electronic structure calculations for the existing, related perovskite superconductors -- MgCNi$_3$ and YRh$_3$B, applying the experimental values of lattice parameters (in Bohr units): 7.205 (Ref.~\cite{nature1}) and 7.870 (Ref.~\cite{ybrh3}), respectively. Since DOS for MgCNi$_3$ was recently published by many authors (e.g. Refs.~\cite{szajek,dugdale}), we present KKR DOS for the less known YRh$_3$B in Fig.~\ref{fig:ybrh3}.

\begin{table}[htb]
\caption{McMillan--Hopfield parameters for MgCNi$_3$ and YRh$_3$B  (mRy/Bohr$^2$/atom).} \label{tab:comp}
\begin{ruledtabular}
\begin{tabular}{lrc|lr}
Element & {$\eta$}  & &Element & $\eta$ \\ \hline
Ni   & 20.4      & & Rh   &  17.5  \\
Mg   & 0.1       & & Y    &  1.4   \\
C    & 9.1       & & B    &  1.6   \\
\end{tabular}
\end{ruledtabular}
\end{table}

As we can observe from Tab.~\ref{tab:xb} and Tab.~\ref{tab:comp}, values of $\eta_i$ for the transition element (Sc,~Ni,~Rh) are very similar in all cases. It is interesting to underline that the presence of the $2p$--states of light element (B in this case) at the Fermi level in YRh$_3$B is negligible  \cite{ybrh3:el,ybrh3:el2} (see~Fig.~\ref{fig:ybrh3}), unlike in MgCNi$_3$ or Sc$_3X$B. This is the reason of a very small $\eta_B$ parameter, comparing to $\eta_C$ in MgCNi$_3$ and especially $\eta_B$ in Sc$_3X$B. Also the large mass of rhodium ($M\simeq$ 103~u), comparing to scandium ($M\simeq$ 45~u) or nickel ($M\simeq$ 59~u) may be responsible for the low superconducting critical temperature ($\sim$1~K) in this compound.
The comparison of McMillan--Hopfield parameters in presented perovskites additionally favors superconductivity in Sc$_3X$B compounds due to relatively low mass of scandium.

%
%
%
%

\subsection{\label{vacancy}Effect of vacancy on B-site}

When dealing with this family of compounds, one has to face with possible crystallographic
imperfections of the Sc$_3X$B structure. The common problem occurring in intermetallic perovskite borides (and carbides) is related
to not fully occupied boron (carbon) position. Note that various boron or carbon
deficiency was observed in many related perovskites  \cite{schaak}, as well as in MgCNi$_3$.
This effect is very important for superconductivity, since in MgC$_x$Ni$_3$
critical temperature decreases linearly with C concentration~\cite{carbon} and superconductivity
disappeares for $x$~$<$~0.9. The vacancy on B--site is highly plausible in our compounds
and this effect has to be taken into account in the analysis.
This prompted us to perform
the calculations of electronic structure and McMillan--Hopfield parameters in Sc$_3X$B$_{x}$ from
the KKR--CPA method \cite{cpa}.
A~vacancy on boron site was treated as an 'empty sphere'
with $Z$~=~0, and the same $MT$--radius as applied for B atom. In the non--stoichiometric Sc$_3X$B$_x$ the (1b) position $({1\over2},{1\over2},{1\over2})$ is occupied by B atom and a vacancy with probabilities $x$ and 1~--~$x$, respectively.
The electronic structure of such disordered system was calculated using the coherent potential approximation (CPA),
which allows considering any finite concentrations, including impurity states.

\begin{figure}[b]
\includegraphics[width=.49\textwidth]{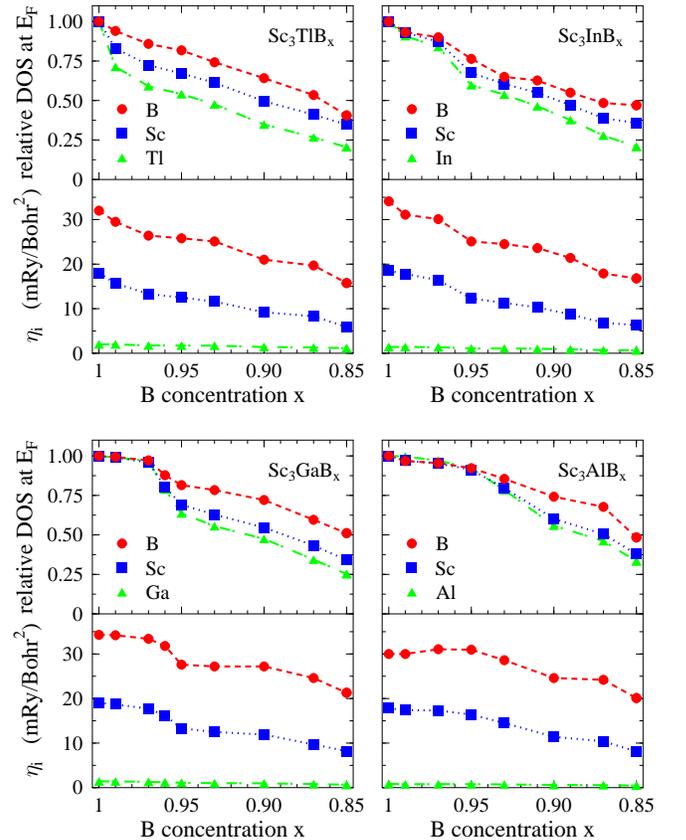}
\caption{\label{fig:vac} Effect of B--sublattice deficiency in Sc$_3X$B$_{x}$ on site--decomposed DOS at the Fermi level (upper panels) and corresponding McMillan--Hopfield parameters (lower panels) computed from KKR--CPA method (see text). In both panels, the corresponding values for B (in red), Sc (in blue) and $X$ (in green) are marked by circles,  squares and triangles, respectively. Lines connecting calculated
points are added as a guide to the eye.}
\end{figure}

Site--decomposed DOS at $E_F$ in Sc$_3X$B$_x$ (divided by $n(E_F)$ in stoichiometric Sc$_3X$B) and McMillan--Hopfield parameters, versus B concentration, are shown in Fig.~\ref{fig:vac}. At a first glance one observes, that all site contributions to DOS at $E_F$ decrease in the investigated systems with decreasing of boron concentration, and reach less than 50\% of initial values for $x$~=~0.85. This is
unlikely to the rigid--band behavior, where $E_F$ is expected to move to the left, towards higher DOS (see~Fig.~\ref{fig:xb}). More detailed analysis of KKR--CPA DOS in Sc$_3X$B$_{x}$ (see~also~Fig.~\ref{fig:cpa}) indicates, that a vacancy on B--site seems to behave as a hole donor, if the pseudo--gap in DOS, found above $E_F$, can be considered as a separation between valence--like and conduction--like bands. Since the potential of the vacancy is much more repulsive than the potential of B atom, all $p$--states (accommodating six electrons) are expelled into higher energy range (well above $E_F$), against only one electron occupying $p$--shell in B atom. Consequently, the filling of low--lying conduction--like states decreases, when the
vacancy concentration increases.

\begin{figure}[t]
\includegraphics[width=.49\textwidth]{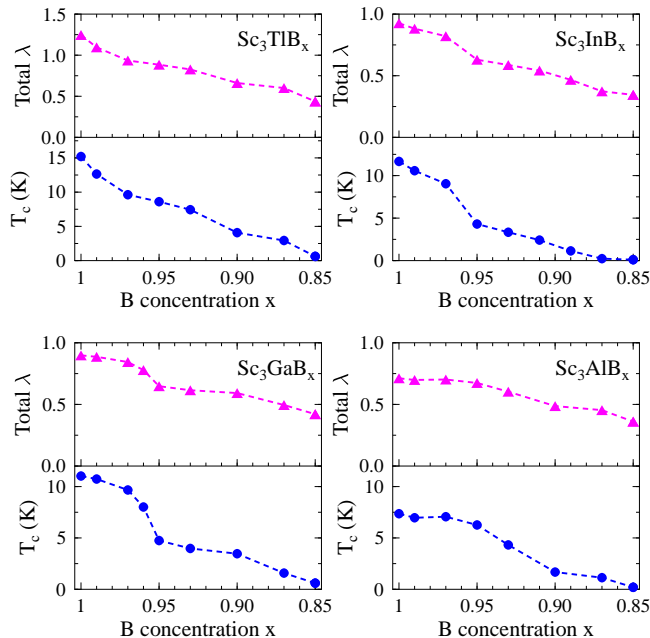}
\caption{\label{fig:vac-tc} Effect of B--sublattice deficiency in Sc$_3X$B$_{x}$ on EPC constant $\lambda$ (upper panels) and corresponding critical temperature (lower panels).}
\end{figure}

The critical parameters analysis for considered vacancy concentrations (0.85~$< x <$~1) shows
that presence of vacancy is very unfavorable for occurrence of superconductivity in these structures, due to
a sudden decrease of the most important $\eta_{Sc}$ and $\eta_{B}$ parameters (Fig.~\ref{fig:vac}). Using the values of $\langle \omega_i^2 \rangle$ obtained in the stoichiometric Sc$_3X$B compounds, the variation of total $\lambda$ and $T_c$ for boron--deficient structures was analyzed. As we can see in Fig.~\ref{fig:vac-tc}, in all structures the decrease in
the EPC constant $\lambda$ is so fast, that when vacancy concentration reaches 15\% (i.e. Sc$_3X$B$_{0.85}$) superconductivity is practically quenched, since $T_c$ $\sim$~0.1~K.
So, the boron atom occurs to play crucial role in superconductivity onset in Sc$_3X$B, and even small boron deficiency may cause rapid decrease of the critical temperature.

This behavior seems to be similar to MgCNi$_3$ case, where lowering of $T_c$ with increase of vacancy concentration on carbon site, was observed experimentally~\cite{carbon,carbon2}.
Quite recently, two important investigations were carried for this system. It was deduced from the specific heat measurements \cite{role_c-spheat} that in a carbon--deficient sample low--energy nickel phonon modes (which are probably the most important in superconductivity) were shifted to higher energies. Furthermore, electronic structure calculations performed for MgC$_x$Ni$_3$ in the range 0.80~$<x<$~1 exhibited the decrease of $n(E_F)$ with the increasing $x$  \cite{role_c-calc}, but less drastic than in our Sc$_3X$B$_x$ systems.
So, disappearance of superconductivity in MgC$_x$Ni$_3$ seems to result from two unfavorable effects: lowering the DOS at Fermi level combined with increasing phonon frequencies. Taking into account the possibility of similar behavior of low--energy phonon modes in Sc$_3X$B$_x$, we may suggest, that superconductivity in Sc$_3X$B could be even more sensitive to stoichiometry, than in MgCNi$_3$.

%
%
%
%

\subsection{\label{magn}Magnetic properties of cubic Sc$_3X$}

The motivation to study relations between superconductivity in Sc$_3X$B, and magnetism in Sc$_3X$ compounds, was inspired by the widely--studied weak itinerant ferromagnetism of the hexagonal Sc$_3$In \cite{matt}.
In spite of recent interest in this field, the research of the cubic form of Sc$_3$In has not been carried out so far. To have a possibility of wider comparison between the two series of compounds: Sc$_3X$B and Sc$_3X$, the electronic structure calculations for three hypothetical structures (Sc$_3$Al, Sc$_3$Ga and Sc$_3$Tl) were also performed.

\begin{table}[t]
\caption{Lattice parameters in Sc$_3X$ series, units: 1~Bohr~=~0.5292~\AA.} \label{tab:lat2}
\begin{ruledtabular}
\begin{tabular}{lcc}
Compound & $a$ experimental \cite{phase_d}  & $a_0$ calculated \\
\hline
Sc$_3$Tl &  ---   &  8.300  \\
Sc$_3$In & 8.427  &  8.150  \\
Sc$_3$Ga &  ---   &  8.025  \\
Sc$_3$Al &  ---   &  8.150  \\
\end{tabular}
\end{ruledtabular}
\end{table}

\begin{figure}[b]
\includegraphics[width=.40\textwidth]{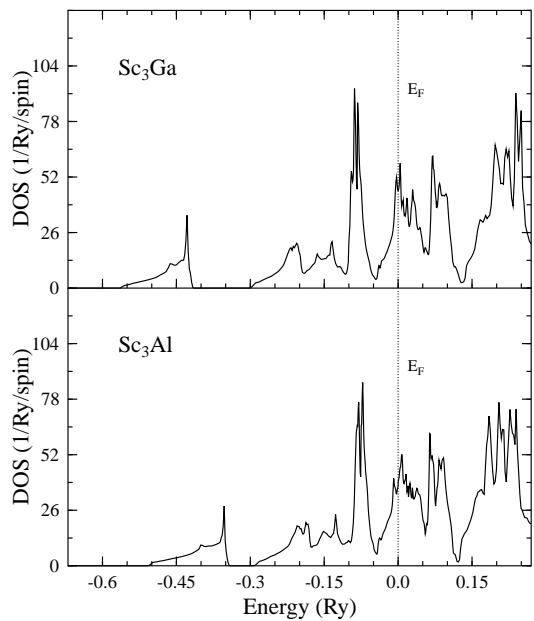}
\caption{KKR non--polarized total DOS in Cu$_3$Au--type Sc$_3$Ga and Sc$_3$Al calculated at equilibrium lattice constant (Tab.~\ref{tab:lat2}). The Fermi level is shifted to zero and marked by a vertical line.}
\label{fig:sc3x-nmag}
\end{figure}

In all cubic Sc$_3X$ structures lattice constants were derived from the total energy calculations, and MT~sphere radii for both Sc and $X$ atoms were set to 0.35$a_0$ (other computational details are the same as in Sec.~\ref{sc}). Table~\ref{tab:lat2}
presents the calculated values of equilibrium lattice parameter ($a_0$). The variation of $a_0$ with $X$--element in  Sc$_3X$ is similar to the tendency observed in Sc$_3X$B. However, there is much larger difference (3\%) between the theoretical and experimental values for Sc$_3$In, but still within the acceptable error of LDA. This however prompted us to study electronic structure as a function of lattice constant in this compound.

First, we have analyzed the non--polarized DOS and the Stoner product for Sc$_3X$ compounds (Tab.~\ref{tab:sc3x}).
In view of the KKR results, the Stoner criterion is not fulfilled only in Sc$_3$Ga and Sc$_3$Al, and indeed, the spin--polarized KKR computations yielded non--magnetic ground state in both cases (non--spin-polarized DOS are
shown in Fig.~\ref{fig:sc3x-nmag}).

\begin{figure}[t]
\includegraphics[width=.38\textwidth]{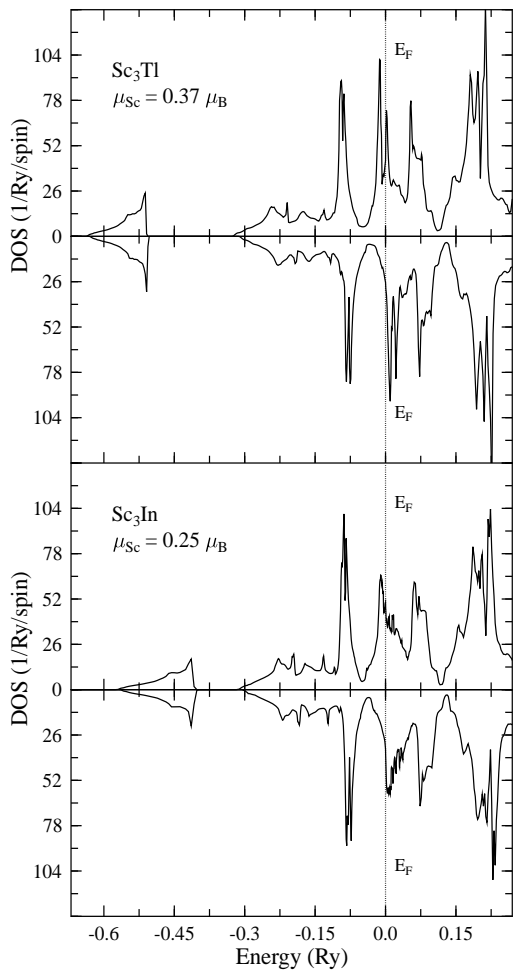}
\caption{KKR polarized total DOS in Cu$_3$Au-type Sc$_3$Tl and Sc$_3$In calculated at equilibrium lattice constant (Tab.~\ref{tab:lat2}). The Fermi level is shifted to zero and marked by a vertical line.}
\label{fig:sc3x-mag}
\end{figure}

\begin{table}[t]
\caption{Electronic properties of Sc$_3X$. $n(E_F)$ is given in (1/Ry/spin/f.u.) (f.u. -- formula unit), $n_{Sc}$~in (1/Ry/spin/atom), $\mu_{Sc}$ in ($\mu_B$/atom).} \label{tab:sc3x}
\begin{ruledtabular}
\begin{tabular}{lrrrrc}
\multicolumn{6}{c}{non--spin--polarized calculations} \\ \hline
Compound & \multicolumn{2}{c}{$n(E_F)$} & \multicolumn{2}{c}{$n_{Sc}(E_F)$}  &  $I\times n(E_F)$ \\
\hline
Sc$_3$Tl &  \multicolumn{2}{c}{91.2} & \multicolumn{2}{c}{30.1}  & 1.78 \\
Sc$_3$In &   \multicolumn{2}{c}{62.1} &  \multicolumn{2}{c}{19.9}  & 1.17 \\
Sc$_3$Ga &   \multicolumn{2}{c}{45.9} &  \multicolumn{2}{c}{14.4}  & 0.89 \\
Sc$_3$Al &   \multicolumn{2}{c}{37.2} &  \multicolumn{2}{c}{11.8}  & 0.70 \\ \hline
Sc$_3$In (hex) &  \multicolumn{2}{c}{72.8}  &   \multicolumn{2}{c}{23.0}  &  1.32\\ \hline \hline
\multicolumn{6}{c}{spin--polarized calculations} \\ \hline
Compound & $n_{\uparrow}(E_F)$ & $n_{\downarrow}(E_F)$ & $n_{\uparrow Sc}(E_F)$ & $n_{\downarrow Sc}(E_F)$ & $\mu_{Sc}$ \\ \hline
Sc$_3$Tl &  43.3 &  26.4 & 14.2 & 8.5 & 0.37  \\
Sc$_3$In &  49.6 &  34.7 & 15.7 & 11.0 & 0.25  \\ \hline
Sc$_3$In (hex)& 69.7 &  27.5 & 19.3 & 8.5 & 0.32   \\
\end{tabular}
\end{ruledtabular}
\end{table}

\begin{figure}[b]
\includegraphics[width=.37\textwidth]{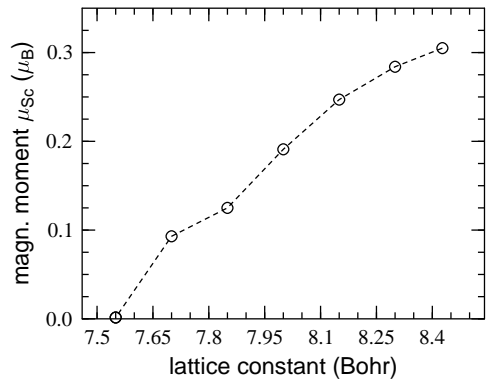}
\caption{Magnetic moment on scandium atom in Sc$_3$In as a function of the lattice constant. Line is added as a guide for the eye.}
\label{fig:mom}
\end{figure}

In agreement with the Stoner analysis, spin--polarized KKR computations in Sc$_3$Tl and Sc$_3$In
converged to ferromagnetic ground state (Fig.~\ref{fig:sc3x-mag}) with magnetic moments as large
as $0.37~\mu_B$ and  $0.25~\mu_B$ per scandium atom for the case of $X$~=~Tl and In, respectively.
The magnetic moment on $X$ atom was negligible ($\simeq$~0.02~$\mu_B$).
Regarding to the calculations, scandium in cubic phase of Sc$_3$In possesses quite similar magnetic moment as in the
hexagonal one~\footnote{In the Ni$_3$Sn--type Sc$_3$In experimental lattice constants were taken: $a$~=~6.42~\AA, $c$~=~5.18~\AA  \cite{matt2}, the adjustable parameter $x$, positioning
Sc atoms in the XY plane, was assumed to have the ideal value of $x=5/6$, following the X--ray
analysis  \cite{matt2} and previous calculations  \cite{singh}. The same MT~sphere radius (3.0~Bohr)
were used for Sc and In. In the present work the magnetic moment is given per Wigner-Seitz cell and per
Sc atom. In Ref.~\cite{pm05} the magnetic moments of 0.26~$\mu_B$ (hexagonal phase) and 0.27~$\mu_B$ (cubic phase, experimental lattice constant), were given per Sc muffin-tin sphere.}: $\mu_{Sc}$~=~0.32~$\mu_B$ (comparing with other LDA result 0.30~$\mu_B$ \cite{singh}).
The magnitude of magnetic moment in Sc$_3$In as a function of lattice constant is shown in Fig~\ref{fig:mom}. The observed monotonic decrease of magnetization with
the cell volume shrinking indicates, that the quantum critical point, when magnetization disappears, can be reached under hydrostatic pressure. Conversely, magnetization in hexagonal Sc$_3$In enhances \cite{sc3in_press}
under moderate hydrostatic pressure, which is supported by electronic structure calculations \cite{singh} suggesting an increase of magnetic moment under weak hydrostatic pressure, while suppression of ferromagnetism upon applying the uniaxial strain.

It seems interesting to recall, that the widely investigated weak ferromagnet Ni$_3$Al \cite{ni3al} crystallizes in the same Cu$_3$Au--type structure. Also, the electronic structure calculations in Ni$_3$Al gave weak magnetic moments on Ni ($\mu_{Ni} \simeq 0.24$~$\mu_B$~\cite{singh_ni}), which is close to the KKR value we have gained for scandium in the cubic Sc$_3$In ($\mu_{Sc} \simeq 0.25$~$\mu_B$). According to measurements \cite{ni3al}, Ni$_3$Al exhibits weak itinerant magnetism, with however much smaller magnetic moment $\mu_{Ni}$ = 0.075~$\mu_B$. Surprisingly, the similar compound Ni$_3$Ga was found to be paramagnetic in experiment \cite{ni3al}, although LDA calculations resulted in the magnetic state with $\mu_{Ni}$ = 0.26~$\mu_B$~\cite{singh_ni}, even larger than in Ni$_3$Al.
The discrepancy between LDA prediction and experimental finding in Ni$_3$Ga was explained in terms of strong
spin fluctuations \cite{singh_ni}.

The question whether Sc$_3X$ systems are similar to Ni$_3X$ compounds will be interesting after experimental verification of existence of these systems.

%
%
%
%

\subsection{\label{sc-fm}Ferromagnetism vs. Superconductivity}

As we can see from the limiting cases, studying possible competition of superconductivity and weak
ferromagnetism is directly related to investigation of the role of trivalent boron in the entitled compounds.
To enlighten this subject, the KKR--CPA calculations were performed in full range of B concentration for
illustrative example of Sc$_3$InB$_x$.

\begin{figure*}[htb]
\includegraphics[width=.95\textwidth]{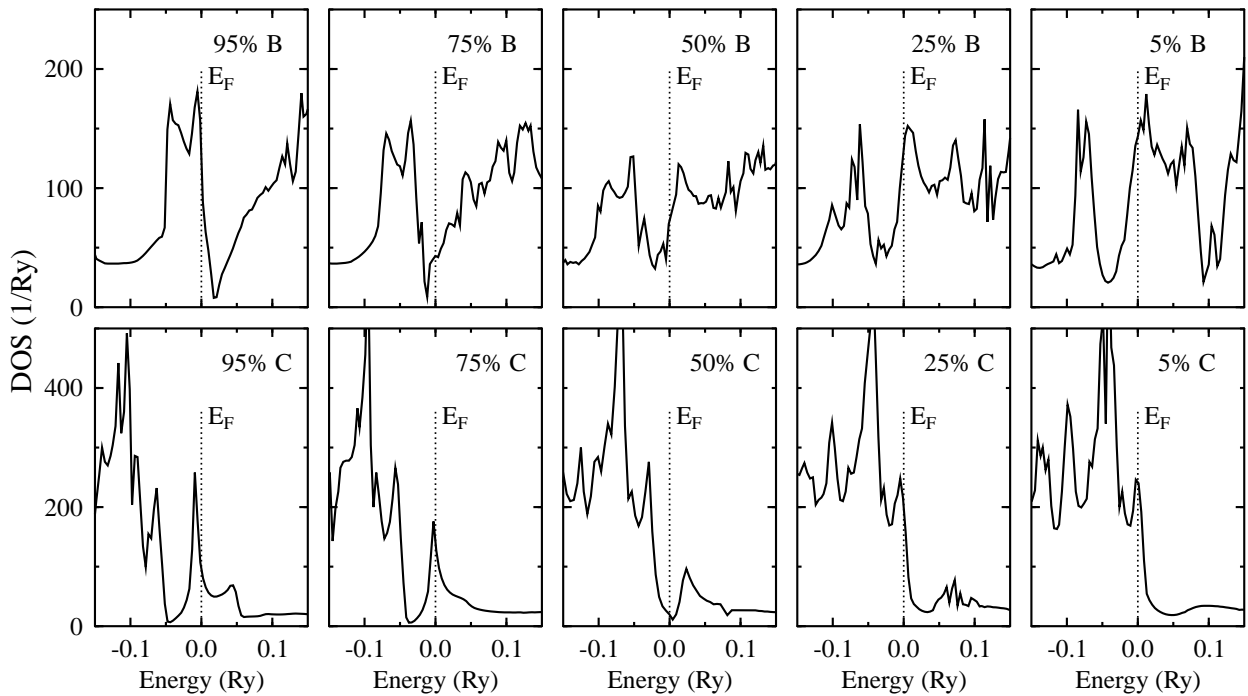}
\caption{\label{fig:cpa} Effect of vacancy on total density of states in Sc$_3$InB$_x$ (upper panels) and   MgC$_x$Ni$_3$ (lower panels) from KKR--CPA calculations. Note a remarkable evolution of electronic states in the vicinity of the Fermi level: from high DOS at E$_F$ in C-rich and B-rich samples (left side: i.e. near full C/B occupancy and superconductivity limit) through a deep DOS minimum (intermediate concentrations, i.e. near 50\% C in MgC$_x$Ni$_3$ and c.a. 80\% B in Sc$_3$InB$_x$) to again large DOS, which satisfy the Stoner criterion (right side, i.e. near empty C/B sublattice limit). The Fermi level is at zero, marked by a vertical line.}
\end{figure*}

Analyzing the non--magnetic DOS values for Sc$_3$In (Tab.~\ref{tab:sc3x}) one notices, that the large $n(E_F)$ value consists mainly (over 95\%) of scandium atoms contribution. This gives large Stoner product on Sc and makes the magnetic ground state energetically favorable.
The DOS evolution (Fig.~\ref{fig:cpa}) shows that when B concentration increases, the Fermi level shifts
from the strongly increasing DOS (5\% B in Fig.~\ref{fig:cpa}), leaving the 'magnetic' region,
towards a deep valley (75\% B in Fig.~\ref{fig:cpa}). After crossing this minimum, the $n(E_F)$ value again increases, but the Stoner limit on Sc
is not reached here (95\% B in Fig.~\ref{fig:cpa}).
The total DOS at $E_F$, as well as the scandium contribution $n_{Sc}(E_F)$ in Sc$_3$InB is much lower, than the corresponding values in Sc$_3$In. Consequently, the Stoner product in Sc$_3$InB is as small as 0.5, which prevents formation of magnetic ground state.

\begin{figure}[t]
\includegraphics[width=.40\textwidth]{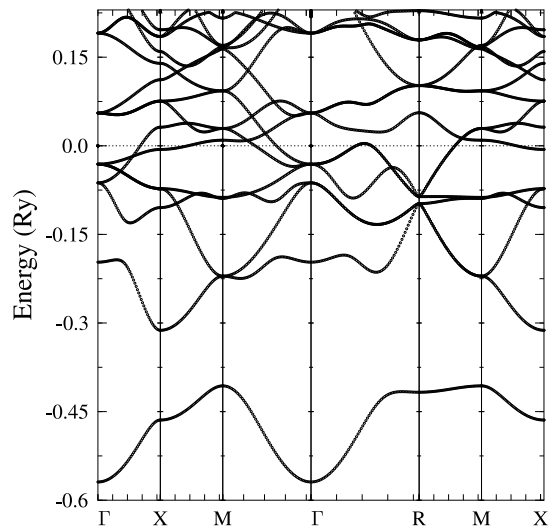}
\caption{\label{fig:bnd2}The non--polarized electronic dispersion curves E({\bf k}) along high symmetry directions
in cubic Sc$_3$In (for $a_0$ = 8.150 Bohr). The Fermi level is marked by a horizontal line.}
\end{figure}

In order to better understand mechanism of lowering the scandium contribution to $n(E_F)$, we can notice
interesting result concerning the formation of one additional low--lying band in Sc$_3$InB
(two lowest--lying bands below -0.3~Ry in Fig.~\ref{fig:bnd}, against only one in Sc$_3$In in
Fig.~\ref{fig:bnd2}). This band is formed from hybridization of $s-$like states from B with Sc orbitals
(including $d-$states) and presumably bounds about one electron from B and one electron from Sc.
Consequently, the energy bands in Sc$_3$InB in the range of $-0.3$~Ry~$<E <E_F$) includes one Sc electron
less, than the bands in Sc$_3$In in the corresponding energy range.

When comparing integrated partial $d$--DOS for Sc atom in both compounds, we clearly notice the
decrease of $d-$Sc orbitals filling in Sc$_3$InB: $1.5~e$ against $1.7~e$ in Sc$_3$In (per Sc atom) as
well as the transfer of about $0.1~e$ from upper bands to the lower additional band for each Sc atom.

These two effects show the important modification of valence states of Sc upon B insertion and seem
also to be responsible for lowering the scandium contribution to $n(E_F)$ and preventing the ferromagnetism.
When the ferromagnetic ground state in Sc$_3$TlB and Sc$_3$InB is destroyed, the conventional superconductivity is allowed to appear.
Certainly, the presence of boron atom is not only important for preventing ferromagnetism, but, as we have seen in Sec.~\ref{vacancy}, it is crucial to promote superconductivity.
This opens the perspectives of interesting experimental study, if it would be possible to synthesize the Sc$_3X$B$_x$ systems with various boron concentration.

Appearance of magnetic ground state in Sc$_3$Tl and Sc$_3$In makes Sc$_3$TlB and Sc$_3$InB even more
similar to MgCNi$_3$, since hypothetical Cu$_3$Au--type structure MgNi$_3$ was theoretically predicted
to have a small magnetic moment (about 0.4~$\mu_B$ per Ni atom \cite{mgni3}). We may confirm this
result ($\mu_{Ni}$ = 0.40 $\mu_B$ from KKR).

For comparison with Sc$_3$InB$_x$ we present the evolution of electronic structure for MgC$_x$Ni$_3$ (Fig.~\ref{fig:cpa}).
The DOS variation with C concentration is a bit different, since $E_F$ moves towards lower lying valence
states when the carbon concentration decreases, in contrast to the Sc$_3$InB$_x$ system. However, the 
general trends are quite similar in both cases. When C concentration decreases the Fermi level
crosses the DOS valley and next falls into the higher--DOS region, where magnetic ground state appears. Also it is worth noting, that van Hove singularity near $E_F$ in MgCNi$_3$, being the unusual feature of
its electronic structure, disappears with the vacancy concentration (due to the deficiency of carbon $p$--orbitals).
Note that our DOS picture for $x$~=~75\% compares well with the result obtained for $x$~=~80\% in Ref.~\cite{role_c-calc}.

Interestingly, the comparison of data presented in Tab.~\ref{tab:tc} and Tab.~\ref{tab:sc3x} suggests, that the strength of \emph{magnetic} interactions in Sc$_3X$ and the \emph{electron--phonon} interactions in Sc$_3X$B seems to be correlated, since both the Stoner product $I\times n(E_F)$ and EPC parameter $\lambda$ are increasing
with $X$, i.e. the lowest values are observed for $X$~=~Al and highest for $X$~=~Tl.

%
%
%
%

\section{Conclusions}

We have presented theoretical investigation of superconducting properties of the perovskite series Sc$_3X$B ($X$~=~Tl,~In,~Ga,~Al) and their possible connections with weak magnetism in the corresponding Sc$_3X$
compounds. Our main results predicted:

\noindent
(i) superconductivity in Sc$_3X$B, with $\lambda \simeq$~0.7 -- 1.25 and
$T_c \simeq$ 7 -- 15~K;

\noindent
(ii) weak ferromagnetism in Sc$_3$In and Sc$_3$Tl and the absence of ferromagnetism in Sc$_3$Ga and Sc$_3$Al;

\noindent
(iii) critical effect of vacancy on B--site on superconductivity in Sc$_3X$B$_x$.

We have also showed that boron inserted to the cubic Sc$_3$Tl and Sc$_3$In destroys the magnetic ground state, and likely turns these systems into superconductors.

On the whole, the estimated EPC parameters and critical temperatures $T_c$ obtained for Sc$_3X$B compounds
from the RMTA analysis are even larger that the values reported for two perovskite superconductors
MgCNi$_3$ and YRh$_3$B.

Preliminary experimental examination of one of the compounds from the entitled series -- Sc$_3$InB -- was undertaken~\cite{pm05}, but several synthesis procedures did not succeed in preparing a single--phase and stoichiometric
compound. However, one of the samples showed signs of superconductivity with $T_c \simeq 4.5$~K.
Further experimental study is needed to clarify this result.

Moreover, in view of the fact that some itinerant ferromagnets exhibit superconducting
properties, e.g. Y$_9$Co$_7$ \cite{akolo} or UGe$_2$ \cite{uge2}, the experimental investigation
of cubic series Sc$_3X$ should also be very appealing.

\end{document}